\documentclass[12pt]{article}
\usepackage[utf8]{inputenc}
\usepackage{amsmath,amssymb}
\usepackage{latexsym}
\usepackage{slashed}
\usepackage{authblk}
\usepackage{xcolor}
\usepackage{hyperref}

\usepackage{rotate}
\usepackage{lscape}
\usepackage{graphicx}
\usepackage{lineno}
\usepackage{caption}
\usepackage{comment}
\usepackage{xspace}

\newcommand{\SANC}{\texttt{SANC}\xspace}

\newcommand{\MCSANCee}{\texttt{MCSANCee}\xspace}
\newcommand{\ReneSANCe}{\texttt{ReneSANCe}\xspace}

\newcommand{\BabaYaga}{\texttt{BabaYaga}\xspace}

\newcommand{\bq}{\begin{equation}}
\newcommand{\eq}{\end{equation}}
\newcommand{\ba}{\begin{eqnarray}}
\newcommand{\ea}{\end{eqnarray}}
\newcommand{\bqa}{\begin{eqnarray}}
\newcommand{\eqa}{\end{eqnarray}}

\def\mev{{\hbox{MeV}}}
\def\gev{{\hbox{GeV}}}

\newcommand{\sss}[1]{\scriptscriptstyle{#1}}

\begin{document} 

\title{Effects of electroweak radiative corrections in polarized low-energy electron-positron 
annihilation into lepton pairs}

\author[1]{A.\,Arbuzov}
\author[1]{S.\,Bondarenko}
\author[2,3]{Ya.\,Dydyshka}
\author[2]{L.\,Kalinovskaya}
\author[2]{L.\,Rumyantsev}
\author[2]{R.\,Sadykov}
\author[2,3]{V.\,Yermolchyk}
\author[2,3]{U.\,Yermolchyk}

\affil[1]{\small Bogoliubov Laboratory of Theoretical Physics, JINR, 
                 141980 Dubna, Moscow region, Russia}
\affil[2]{\small Dzhelepov Laboratory of Nuclear Problems, JINR,  
                 141980 Dubna, Moscow region, Russia}
\affil[3]{\small Institute for Nuclear Problems, Belarusian State University,
  Minsk, 220006  Belarus}

\maketitle

\begin{abstract}
Complete one-loop electroweak radiative corrections 
to the cross section of the process 
$e^+e^- \to \mu^-\mu^+ (\tau^-\tau^+)$
are evaluated with the help of the \SANC system.
Higher-order contributions of the initial state radiation are computed in the QED structure function formalism.
Numerical results are given for the center-of-mass energy range $\sqrt{s}=5, ~7$~GeV
for various polarization degrees of the initial particles.
\end{abstract}


\section{Introduction}

Processes of electron-positron annihilation provide a powerful tool 
in studies of elementary particles. In particular, modern $e^+e^-$ colliders,
such as VEPP-2000 (Novosibirsk), BEPC II (Beijing), KEKB (Tsukuba) etc., 
are well suited for production and high-precision studies of hadrons.
Electron-positron colliders have significant advantages: clean signals, a low background, a high efficiency 
and resolution. The continuously increasing experimental accuracy challenges
the theory to provide more and more precise predictions. 
For example, the current and upcoming experiments
SuperKEKB~\cite{Akai:2018mbz}, BES-III~\cite{Asner:2008nq},
Super Charm-Tau Factory~\cite{Epifanov:2020elk} and Super Tau-Charm Facility~\cite{Peng:2020orp}
aim at reaching an error of a few per mille in luminosity measurements.
This requires new calculations with taking into account higher order perturbative
corrections and other effects including electroweak ones. 

Another important advantage of $e^+e^-$ colliders is the possibility of using
polarized beams. Several future projects of such machines foresee having at least
longitudinally polarized electron beams. That will open new possibilities
in high-precision studies of the charm quark and tau lepton physics.
A very high accuracy of checking the universality 
of the neutral current vector couplings and in searches for
CP-violation in the lepton sector will be achieved.  
A new independent measurement of the effective electroweak mixing parameter 
$\sin^2\theta_W$ through left-right asymmetries will be complementary to 
the corresponding studies at higher energies. 
Measurements with polarized beams will also help to
refine the elements of the Cabibbo-Kobayashi-Maskawa matrix,
study QCD at low energies and exotic hadrons, search for new physics and extensively investigate two-photon physics.
The SuperKEKB team (Belle collaboration) ~\cite{Belle-II:2018jsg} considers plans 
of an upgrade to have longitudinally polarized electron beam~\cite{Roney:2021Bd,liptak:ipac2021-thpab022}.  
That will significantly widen the collider's capability of examining the electroweak sector. 

BES-III has collected more than 35 fb$^{-1}$ of integrated luminosity at different 
center-of-mass system (c.m.s.) energies from 2.0 to 4.94~GeV.  The upgrade of BES-III will increase the peak 
luminosity by a factor of 3 for beam energies from 2.0 to 2.8 GeV (CM energies from 4.0 to 5.6 GeV).
Future Super Charm-Tau Factories 
 (Super Charm-Tau Factory project~\cite{Epifanov:2020elk} and  
 High Intensity Electron Positron Advanced Facility (HIEPAF)~\cite{Peng:2020orp})
are accelerator complexes
for high-precision measurements between 2 and 5(7) GeV
with luminosity up to $10^{35} cm^{-2} s^{-1}$ and longitudinal
polarization. They will deliver up to 1~ab$^{-1}$ of integrated luminosity per year.

In connection with these challenges, one of the most demanded processes 
is the lepton pair production (LPP) both for
estimating luminosity and for physics program.
Various experimental facilities operating at low energies are in plans or 
already in action demanding appropriate software for theoretical predictions.
At the moment, the most advanced and widely used generators with one-loop radiative 
corrections (RC) for estimation of LPP at low energies
are
\BabaYaga~\cite{CarloniCalame:2000pz,CarloniCalame:2001ny,Balossini:2006wc,Balossini:2008xr},
{\tt KKMC}~\cite{Jadach:1999vf,Jadach:2013aha},
and
{\tt MCJPG}~\cite{Arbuzov:2005pt}.
Recently, a new Monte-Carlo generator~\cite{Nugent:2022ayu}
for the simulation of lepton pair productions
and $\tau$
lepton decays up to an energy of about 11~GeV has been presented.

The \SANC Monte Carlo event generator \ReneSANCe~\cite{Sadykov:2020any} and integrator \MCSANCee 
are relatively new software tools. They can be used in the mentioned energy domain
of electron--positron colliders for simulation of LPP and Bhabha processes.
These tools provide the possibility of evaluating the
complete one-loop QED and (electro)weak radiative corrections. 
Some higher-order leading corrections are also implemented.
In addition, the tools produce results in the full phase space and also
allow taking into account longitudinal beam polarizations. 
To match the high precision of current and near-future experiments we plan to implement
also higher-order next-to-leading QED radiative corrections.
 
In this article, we analyze the effects due to electroweak RC and polarization of the colliding beams using the \SANC software.
We consider the processes of electron-positron annihilation into a lepton pair
\begin{eqnarray}
  e^+(p_1, \chi_1) + e^-(p_2, \chi_2) \to
  \nonumber\\
  l^-(p_3, \chi_3) 
            + {l}^+(p_4, \chi_4) (+ {\gamma}(p_5, \chi_5)),
\label{LPP} 
\end{eqnarray}
where $l=\mu,\tau$ with allowance for arbitrary longitudinal polarization of the initial particles 
($\chi_i$ correspond to the helicities of the particles).
We keep in mind experiments at relatively low c.m.s. energies 
up to about 7~GeV which is relevant for the Super Charm-Tau Factory.
Our aim is to analyze the size of different RC contributions, 
estimate the resulting theoretical uncertainty, and verify the necessity to
include other higher-order corrections.
 
The article is organized as follows.
In Section 2, we discuss various
contributions to the cross sections.
In Section 3, the corresponding numerical results
are  given for the total and differential cross sections.
We consider in detail all possible contributions 
to the cross sections at c.m.s. energies of $\sqrt{s}=5$ and $7$~GeV.
Numerical results are obtained by an estimate of polarization effects.
In Section 4, we analyze the results.

\section{The state-of-the-art radiative corrections at low energies in \SANC}

For a detailed analysis, we divide the contributions to the full correction into several 
parts: the Born level cross section, electroweak (EW) corrections, 
contribution from vacuum polarization, and multiple photon emission effects.

\underline{Born level}

We evaluate the Born level cross section (leading order, LO) contribution 
for two cases: 1) with pure photon exchange $\sigma^{\rm Born}_{\rm QED}(\gamma)$ and 
2) with both photon and $Z$ boson exchange $\sigma^{\rm Born}_{\rm QED}(\gamma, Z)$. 

\underline{Electroweak corrections}

We have already described in detail the technique and results of the analytic calculations 
of the scalar form factors and helicity amplitudes of the general LPP process~(\ref{LPP}) in
our recent paper~\cite{Bondarenko:2020hhn}.
For EW corrections, we calculate the following contributions 
and introduce the notation for them: 

\vspace{3mm}

$\bullet${~QED level}
\vspace{3mm}

Gauge invariant subsets of QED corrections are evaluated separately, 
i.e., the initial state radiation (ISR), 
the final state radiation (FSR), 
and the initial-final interference (IFI).

\vspace{3mm}

$\bullet${~Weak and higher order corrections}

\vspace{3mm}

At low energies, weak-interaction contributions are typically small
since they are suppressed by the ratio $s/M_Z^2$. But for high-precision
measurements they might be still numerically relevant.
We have found it appropriate to combine the 
contributions of the same order of smallness, i.e.,  
weak and higher-order corrections. The corresponding relative
contributions will be further denoted as 
$\delta^{\rm weak}$ and $\delta^{\rm ho}$.
We also distinguish here two possibilities: 
1) the complete one-loop $\delta^{\rm weak}$, where
pure weak-interaction and vacuum polarization (VP) contributions are
taken into account\footnote{
It is conventional in SANC to include the VP contribution into the weak subset of corrections.
}; and 
2) the pure weak-interaction contribution
$\delta^{\rm weak - VP} = \delta^{\rm weak} - \delta^{\rm VP}$.

We evaluate the leading higher-order EW corrections $\delta^{\rm ho}$ 
to four-fermion processes through the $\Delta\alpha$ and $\Delta\rho$ parameters.
A detailed description of our implementation of this contribution was 
presented in~\cite{Arbuzov:2021oxs}.  

\underline{Vacuum polarization}

We introduce two options 
to account for the contribution of vacuum polarization:
$\delta^{\rm VP}_1$ is the choice of hadronic vacuum polarization $\Delta \alpha^{(5)}_{had}(M_{\sss Z})$ part using a parametrization with auxiliary quarks masses, and
$\delta^{\rm VP}_2$ is the choice
using public versions of the {\tt AlphaQED} code by F.~Jegerlehner~\cite{Jegerlehner:2017zsb}.

\underline{Multiple photon effects}

The implementation into \SANC of the multiple photon effects, i.e. ISR (FSR) corrections in
the leading logarithmic approximation (LLA) through the apparatus of QED structure 
functions~\cite{Kuraev:1985hb,Nicrosini:1986sm}
which was described in detail in~\cite{Arbuzov:2021zjc}.
Results are shown up to $\mathcal{O}(\alpha^3L^3)$
finite terms for an exponentiated representation and up to $\mathcal{O}(\alpha^4L^4)$ 
for order-by-order calculations.
The corresponding relative corrections are denoted below as
$\delta^{\rm LLA,ISR(FSR,IFI)}$. 

Particular contributions to ISR(FSR) are sensitive to experimental cuts.
Our cuts are appropriate to the conditions of the Super Charm-Tau Factory project.

The master formula for a general $e^+e^-$ annihilation cross section
with ISR QED corrections in the leading logarithmic approximation
has the same structure as the one for the Drell-Yan process.
For ISR corrections in the annihilation channel, the large 
logarithm is $L=\ln({s}/{m_e^2})$ where the total c.m.s.
energy $\sqrt{s}$ is chosen as the factorization scale.

In the LLA approximation we separate the pure photonic corrections
(marked ``$\gamma$'') 
and the remaining ones which include the pure pair and mixed photon-pair
effects (marked ``$e^+e^-$'' or ``$\mu^+\mu^-$''). Here we do not consider the correction due to light hadron pairs. Numerically, it is comparable with the muon pair contribution but strongly depends on the event selection procedure. So, the corrections due to hadronic effects will be treated elsewhere. 
The corresponding relative corrections are denoted as $\delta^{\rm LLA,i}(k)$ 
with $i=$ ${\rm ISR,FSR,IFI}$, and $k$ suggests the correction type: $\gamma$, 
$e^+e^-$ pairs, or $\mu^+\mu^-$ pairs.

\section{Numerical results and comparisons}
\label{sec2}

In this section, we show numerical results for electroweak radiative
corrections to the annihilation process~(\ref{LPP}) 
obtained by means of the \SANC system.
Numerical results contain estimates of polarization effects.
We compute total cross sections as well as angular distributions 
at the one-loop level.

Here we used the following  set of input parameters:
\begin{eqnarray}
\alpha^{-1}(0) &=& 137.035999084,
\\
M_W &=& 80.379 \; \gev, \quad M_Z = 91.1876 \; \gev,
\nonumber\\
\Gamma_Z &=& 2.4952 \; \gev, \quad m_e = 0.51099895000 \; \mev,
\nonumber\\
m_\mu &=& 0.1056583745 \; \gev, \quad m_\tau = 1.77686 \; \gev,
\nonumber\\
m_d &=& 0.083 \; \gev, \quad m_s = 0.215 \; \gev,
\nonumber\\
m_b &=& 4.7 \; \gev, \quad m_u = 0.062 \; \gev,
\nonumber\\
m_c &=& 1.5 \; \gev, \quad m_t = 172.76 \; \gev.
\nonumber
\end{eqnarray}

The invariant mass and angular cuts are applied to the final state leptons:
\begin{eqnarray}
  \label{cuts}
  |\cos{\theta_{\mu^-}}| < 0.9, ~|\cos{\theta_{\mu^+}}| < 0.9,~M_{l^+l^-} \ge 1\; \gev,
\end{eqnarray}
where $\theta_{\mu^{\pm}}$ are angles with respect to the beam axis.

All calculations are done in the $\alpha(0)$ EW scheme in order
to have a direct access to the effect of vacuum polarization. 
In this scheme, the fine structure constant $\alpha(0)$ and all particle masses 
are input parameters.
All the results are obtained for the c.m.s. energies $\sqrt{s}=5$ and 7~GeV
and for the following three sets of magnitudes of the electron $(P_{e^-})$ and positron $(P_{e^+})$
beam polarizations:
\bqa \label{SetPolarization1}
&& (P_{e^-}, P_{e^{+}}) =
(0,0),(-0.8,0),(0.8,0).
\eqa

\subsection{Different radiative correction contributions}

In order to quantify the impact of different contributions,
we divide them into several parts: three gauge-invariant subsets of QED one-loop corrections, 
the vacuum polarization 
contribution, the weak interaction effects, and the higher order LLA QED contributions. The three QED RC subsets are due to the
initial state radiation (ISR), the final state radiation (FSR), and the interference of the initial and final state radiation (IFI).

The corresponding results for the total LPP cross section are presented in Table~\ref{Table:delta_250plus},
where the relative corrections $\delta^i$ are computed as the ratios (in percent) of the corresponding 
RC contributions to the Born level cross section. Table~\ref{Table:LLAQED5} illustrates the size 
of the ISR higher-order QED corrections computed within the collinear leading logarithmic approximation. 
\begin{table}[ht]
\caption{\label{Table:delta_250plus}Integrated Born and one-loop cross sections and
relative corrections 
for the $e^+ e^- \to \mu^- \mu^+ (\gamma)$ process
at the c.m.s. energies $\sqrt{s}=5$ and $7$~GeV.}
\begin{center}
\begin{tabular}{|l|c|c|}
\hline
$\sqrt{s}$, GeV                & 5           & 7 \\
\hline 
$\sigma^{\rm Born}$, pb        & 2978.58(1)  & 1519.55(1)
\\
$\delta^{\rm weak-VP}$, \%     & 0.029(1)   & 0.005(1) 
\\
$\delta^{\rm VP}_1$, \%          & 5.467(1)   & 6.272(1)
\\
$\delta^{\rm VP}_2$, \% & 5.430(1) & 6.250(1)
\\
$\delta^{\rm ho}$, \%          & 0.224(1)   & 0.294(1)
\\
$\delta^{\rm QED,\ ISR}$, \%    & 8.455(2)  & 9.063(1)
\\
$\delta^{\rm QED,\ FSR}$, \%    & -0.016(1)  & -0.014(1)
\\
$\delta^{\rm QED,\ IFI}$, \%    & 0.012(2)  & 0.017(1)
\\
$\delta^{\rm LLA,\ ISR}$, \%    &  0.668(1) & 0.850(1)
\\
$\delta^{\rm LLA,\ FSR}$, \%    & 0.047(1) & 0.070(1)
\\
\hline
\end{tabular}
\end{center}
\end{table}

\begin{table}[ht]
\caption{\label{Table:LLAQED5}Higher-order ISR corrections in the LLA approximation
for the $e^+e^- \to \mu^- \mu^+ (n\gamma)$ process at $\sqrt{s} = 5$ 
and $7$ GeV. 
Here $\delta_{\text{\tiny{ISR LLA}}} \equiv \delta\sigma_{\text{\tiny{ISR LLA}}}/\sigma_{0}\times 100\%$.}
\centering
\begin{tabular}{|l|c|c|}
\hline
 &
\multicolumn{2}{c|}{$\delta, \%$}\\
\hline
$\sqrt{s}$, GeV                & 5           & 7 \\
\hline
$\mathcal{O}(\alpha^2L^2)$, $\gamma$ & 0.315(1)& 0.436(1) \\
$\mathcal{O}(\alpha^2L^2)$, $e^+e^-$ & 0.238(1)& 0.258(1) \\
$\mathcal{O}(\alpha^2L^2)$, $\mu^+\mu^-$ & 0.100(1)& 0.114(1) \\
$\mathcal{O}(\alpha^3L^3)$, $\gamma$ & -0.008(1)& -0.004(1) \\
$\mathcal{O}(\alpha^3L^3)$, $e^+e^-$ & 0.016(1)& 0.033(1) \\
$\mathcal{O}(\alpha^3L^3)$, $\mu^+\mu^-$ & 0.007(1)&  0.015(1)\\
\hline
\end{tabular}
\end{table}

\begin{figure}[ht]
\begin{center}
\includegraphics[width=0.5\textwidth]{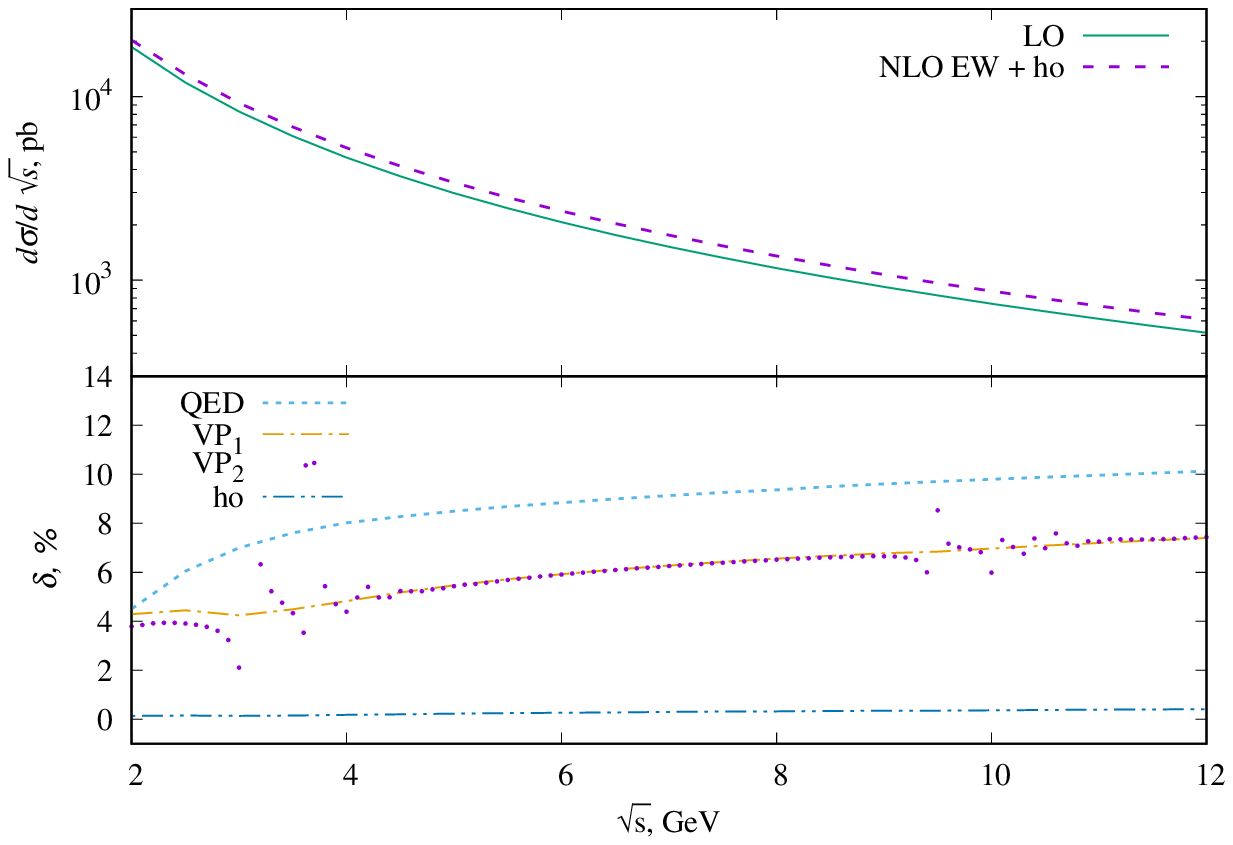}
\end{center}
\caption{
\label{fig:eemumudelta}
Born level (LO) and corrected (EW NLO + ho) cross sections of the $e^+ e^- \to \mu^- \mu^+ (\gamma)$ 
process for the c.m.s. energy range $\sqrt{s}$ = 2 -- 12 GeV (upper panel). 
The relative corrections of the QED, vacuum polarization (VP), and higher-order (ho) contributions (lower panel).}
\end{figure}

In the upper panel of Figure~\ref{fig:eemumudelta}, the Born cross section and the corrected one
which includes the EW NLO and higher-order corrections are shown as a function of the 
initial c.m.s. energy ($\sqrt{s}$ = 2 -- 12~GeV) of the electron and positron 
beams. The cross sections show the fast drop from the $\sim 20$ nanobarn at the c.m.s energy $\sqrt{s} = 2$~GeV 
to about one-half of a nanobarn at $\sqrt{s} = 12$~GeV.

In the lower panel of Figure~\ref{fig:eemumudelta} the relative corrections to the Born cross section are shown in parts, 
namely, the QED, vacuum polarization (VP) and EW higher-order contributions.
The main impact is due to the QED effects, being from 5\% to 10 \%.
The vacuum polarization contribution is also large, ranging from 4\% to 8\%. 
The contribution of higher order is proportional to $\alpha^n$ $(n\geq 2)$ and makes up about 
0.1 -- 0.4 \% because of being enhanced by large logarithms.
Vacuum polarization is treated in two ways. In the first one, the hadronic part of 
vacuum polarization is parameterized by auxiliary quarks masses (it is marked ``VP$_1$'' on the plot). 
In the second one, we use the parameterization by F.~Jegerlehner which accounts for hadron 
resonances~\cite{Jegerlehner:2017zsb} (it is marked ``VP$_2$'' on the plot). 
In both cases, leptonic contributions are taken into account. One can see that the two 
parameterizations agree well at higher energies in regions without resonance. 
But in general, the second parameterization is more appropriate
for the given energy range.

\section{Comparison with \BabaYaga code}

\begin{table}[ht]
    \caption{Tuned comparison of the Born and QED NLO
    integrated cross sections produced by the \SANC and \BabaYaga codes\label{tab:sancvsbabayaga}}
    \centering
    \begin{tabular}{|c|l|l|}
    \hline
    $\sqrt{s}$, GeV		& 5		&    7 \\
    \hline
    \multicolumn{3}{|c|}{Born, nb}\\
    \hline
    \SANC ($Z/\gamma$)    &    2.9786(1) &  1.5195(1)\\
    \SANC (only $\gamma$) &    2.9786(1) &  1.5196(1)\\
    \BabaYaga             &    2.9786(1) &  1.5196(1)\\
    \hline
    \multicolumn{3}{|c|}{QED NLO, nb}\\
    \hline
    \SANC ($Z/\gamma$)    &    3.2304(1)  & 1.6575(1)\\
    \SANC (only $\gamma$) &    3.2287(1)  & 1.6565(1)\\
    \BabaYaga             &    3.2285(1)  & 1.6565(1)\\
    \hline
    \end{tabular}
\end{table}

In Table~\ref{tab:sancvsbabayaga}, we present a tuned comparison of the Born and QED NLO
(without VP contribution) integrated cross sections produced by the \SANC and \BabaYaga codes. 
The results are obtained for two c.m.s. energies $\sqrt{s} = 5$ and 7~GeV
with the angular cuts ~(\ref{cuts}). Very good agreement within statistical errors 
of the results produced by two codes is found.

\begin{figure}[ht]
\centering
\includegraphics[width=0.5\textwidth]{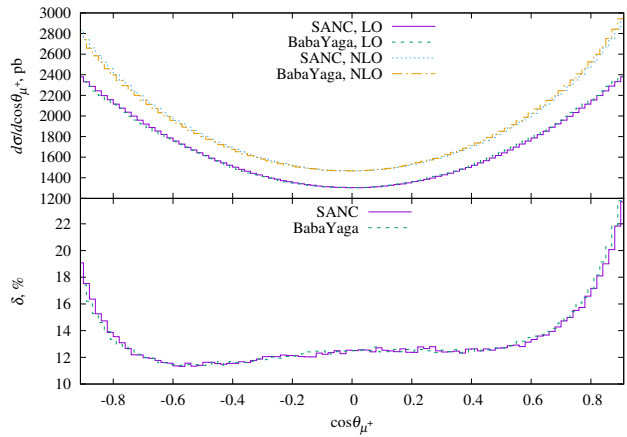}
\caption{
\label{fig:compare-5gev}
LO and NLO unpolarized QED cross sections (upper panel) 
and the relative corrections (lower panel) produced by the \SANC and \BabaYaga codes
for the c.m.s. energy $\sqrt{s}=5$~GeV
as a function of $\cos\theta_{\mu^-}$.}
\end{figure}

\begin{figure}[ht]
\centering
\includegraphics[width=0.5\textwidth]{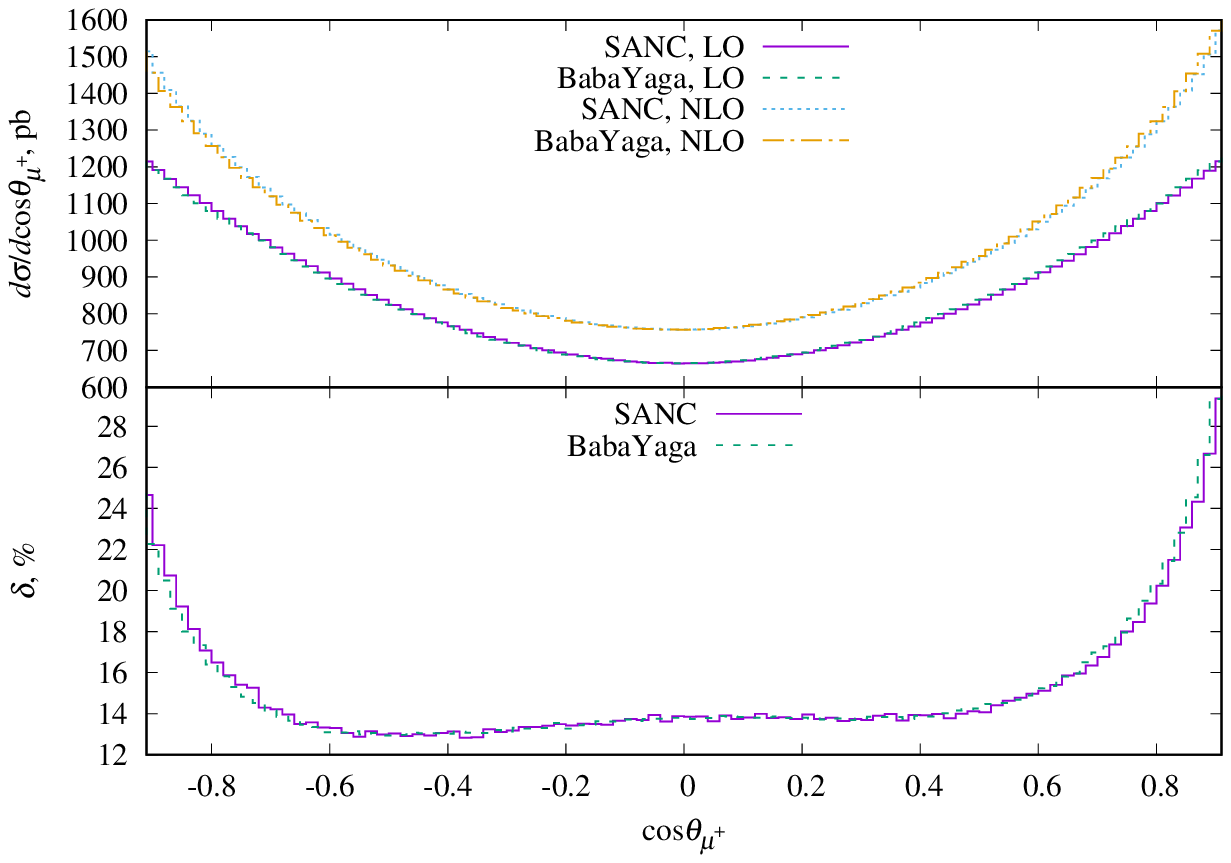}
\caption{
\label{fig:compare-7gev}
Same as in Figure~\ref{fig:compare-5gev} but for c.m.s. energy $\sqrt{s}=7$~GeV.}
\end{figure}

Figures~\ref{fig:compare-5gev} and \ref{fig:compare-7gev} show the comparison of the
\SANC and \BabaYaga results produced for two c.m.s. energies $\sqrt{s}$ = 5 and 7~GeV. 
In the upper panels of the figures, the LO and QED NLO differential cross sections as functions 
of the cosine of the outgoing muon momentum angle are shown.
In the lower panels the relative corrections are compared. 
For both c.m.s energies,
cross sections and the relative corrections are in good agreement. 
To obtain the results for Table~\ref{tab:sancvsbabayaga} and 
Figures~\ref{fig:compare-5gev},\ref{fig:compare-7gev}, additional
extra efforts were made to exclude the $Z$ boson exchange contribution
in the LO and NLO cross sections in the \SANC code.

\subsection{Polarization dependence of cross sections}

Tables \ref{eemumu5pol}-\ref{eetata5pol}
present the integrated Born and one-loop cross sections in pb
and relative corrections in percent for the process $e^+e^- \to l^-l^+$  
at the c.m.s. energy of 5~GeV and set~(\ref{SetPolarization1}) of the initial particle degree of
polarization in the $\alpha(0)$ EW scheme.
The cases with the 7~GeV c.m.s. energy for the $\mu^+\mu^-$
and $\tau^+\tau^-$ final states are presented in 
Tables~\ref{eemumu7pol} and \ref{eetata7pol}, respectively.
It is interesting that the Born and corrected cross sections do 
depend on beam polarizations while the relative correction is almost constant.

\begin{table}[ht]
\caption{
	\label{eemumu5pol}
Polarized integrated Born cross section
and relative corrections
for the $e^+ e^- \to \mu^- \mu^+ (\gamma$) scattering
for $\sqrt{s} = 5$~GeV
for different degree of polarization of the initial particles.}  
    \begin{center}
	\begin{tabular}{|r|l|l|l|}
	    \hline
	    {$P_{e^+}$},{$P_{e^-}$} & $\sigma^{\text{Born}}$, pb & {$\sigma^{\text{1-loop}}$, pb} & {$\delta$, \%}\\
	    \hline
	    $0$, $0$     & 2978.6(1) & 3434.2(1) & 15.30(1)\\
	    \hline
	    $0$, $+0.8$  & 2979.1(1) & 3434.6(1) & 15.29(1)\\
	    \hline	    
	    $0$, $-0.8$  & 2978.0(1) &  3433.7(1) & 15.30(1)\\
	    \hline
    \end{tabular}
    \end{center}
\end{table}

\begin{table}[ht]
\caption{\label{eemumu7pol}
Same as in Table ~\ref{eemumu5pol}, 
but for $\sqrt{s} = 7$~GeV.
	}  
    \begin{center}
	\begin{tabular}{|r|l|l|l|}
	    \hline
	    {$P_{e^+}$},{$P_{e^-}$} & $\sigma^{\text{Born}}$, pb & {$\sigma^{\text{1-loop}}$, pb} & {$\delta$, \%}\\
	    \hline
	    $0$, $0$     & 1519.6(1) & 1773.8(1) & 16.73(1)\\
	    \hline
	    $0$, $+0.8$  & 1520.1(1) & 1774.1(1) & 16.71(1)\\
	    \hline	    
	    $0$, $-0.8$  & 1519.0(1) & 1773.6(1) & 16.76(1)\\
	    \hline
    \end{tabular}
    \end{center}
\end{table}

\begin{table}[ht]
\caption{
\label{eetata5pol} Polarized integrated Born cross section
and relative corrections
for $e^+ e^- \to \tau^- \tau^+ (\gamma)$ scattering
for $\sqrt{s} = 5$~GeV.
}  
    \begin{center}
	\begin{tabular}{|r|l|l|l|}
	    \hline
	    {$P_{e^+}$},{$P_{e^-}$} & $\sigma^{\text{Born}}$, pb & {$\sigma^{\text{1-loop}}$, pb} & {$\delta$, \%}\\
	    \hline
	    $0$, $0$     & 2703.3(1) & 2816.7(1) & 4.20(1)\\
	    \hline
	    $0$, $+0.8$  & 2703.8(1) & 2816.9(1) & 4.18(1)\\
	    \hline	    
	    $0$, $-0.8$  & 2702.8(1) & 2816.5(1) & 4.21(1)\\
	    \hline
    \end{tabular}
    \end{center}
\end{table}

\begin{table}[ht]
    \caption{
  \label{eetata7pol}  
Same as in Table ~\ref{eetata5pol} but for $\sqrt{s} = 7$~GeV.
}  
    \begin{center}
	\begin{tabular}{|r|l|l|l|}
	    \hline
	    {$P_{e^+}$},{$P_{e^-}$} & $\sigma^{\text{Born}}$, pb & {$\sigma^{\text{1-loop}}$, pb} & {$\delta$, \%}\\
	    \hline
	    $0$, $0$     & 1503.0(1) & 1648.8(1) & 9.70(1)\\
	    \hline
	    $0$, $+0.8$  & 1503.6(1) & 1649.1(1) & 9.68(1)\\
	    \hline	    
	    $0$, $-0.8$  & 1502.4(1) & 1648.5(1) & 9.72(1)\\
	    \hline
    \end{tabular}
    \end{center}
\end{table}

\section{Forward-backward asymmetry}
 \label{Sect:AFB}

The forward-backward asymmetry  $A_{\texttt FB}$ is defined as
\bqa
&& A_{\rm FB} = \frac{\sigma_{\rm F}-\sigma_{\rm B}}{\sigma_{\rm F}+\sigma_{\rm B}},
\nonumber \\
&& \sigma_{\rm F} = \int\limits_0^1 \frac{d\sigma}{d\cos\vartheta_f}d\cos\vartheta_f,
\qquad
\sigma_{\rm B} = \int\limits_{-1}^0 \frac{d\sigma}{d\cos\vartheta_f}d\cos\vartheta_f,
\eqa
where $\vartheta_f$ is the angle between the momenta of the incoming electron and the outgoing
negatively charged fermion.
It can be measured in any $e^+e^- \to f \bar{f}$ channels but
for precision tests the most convenient channels are $f=e,\mu$.

\begin{figure}[ht]
\centering
\includegraphics[width=0.5\textwidth]{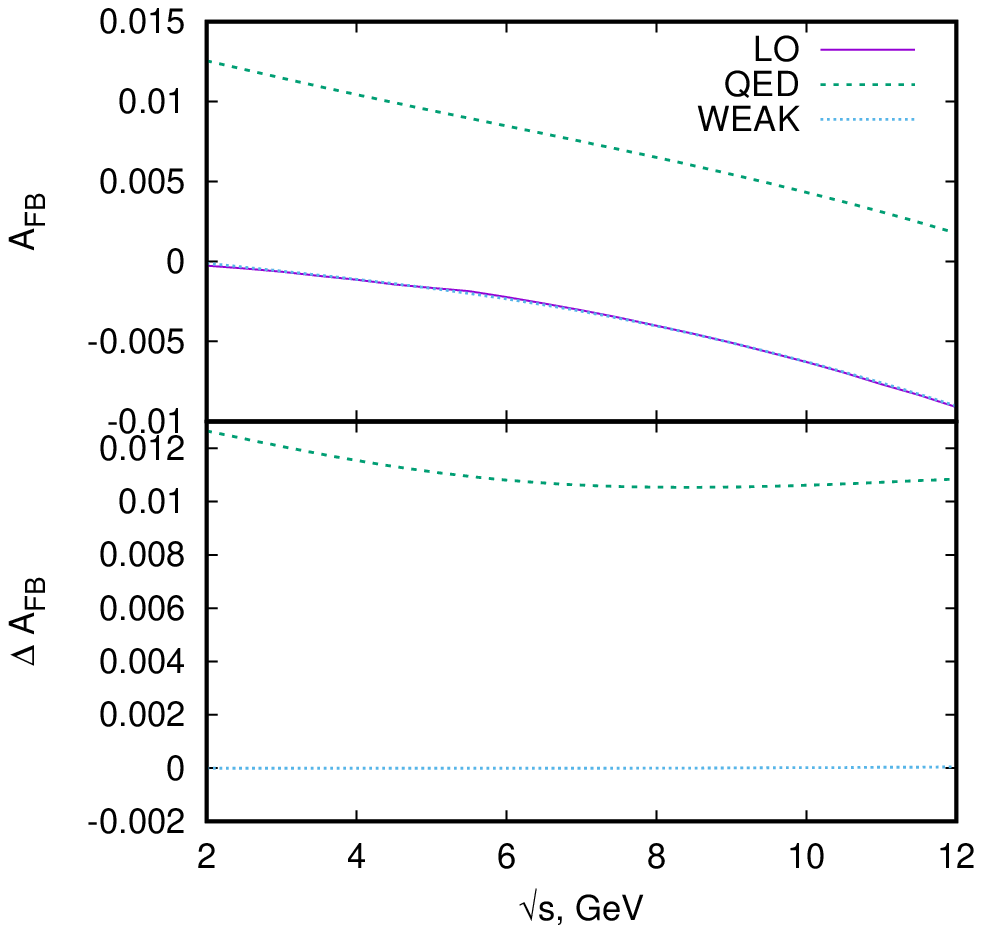}
\caption{
\label{figafb}
  $A_{\rm{FB}}$  asymmetry in the Born and one-loop approximations
  and the corresponding shifts $\Delta A_{\texttt{FB}}$ for the c.m.s. energy range $\sqrt{s}$ = 2 -- 12 GeV.}
\end{figure}

In Figure~\ref{figafb} we show the behavior of the $A_{\texttt{FB}}$ asymmetry
in the Born and 1-loop approximations (with weak, pure QED, or complete
EW RC contributions) 
and of the corresponding $\Delta A_{\texttt{FB}}$ for the c.m.s. energy range 
$2 \leq \sqrt{s} \leq 12$~GeV. 
The asymmetry in the lowest-order approximation 
comes from the tree-level $Z$ boson exchange contribution. One can see that for higher energies it is comparable in size with the QED contribution which comes from one-loop radiative corrections. 

\section{Conclusions}
 
In this paper, we considered different contributions of EW corrections to processes 
of electron-positron annihilation into a lepton pair. The corrections were evaluated 
within the {\SANC} system framework in the $\alpha(0)$ EW scheme for c.m.s. energies up 
to about 10~GeV which are relevant for the existing and future meson factories.  
The complete one-loop EW corrections as well as some leading higher-order corrections 
were analyzed. We see that the pure QED and vacuum polarization corrections dominate
in the given energy range, but in some cases the $Z$ boson exchange amplitude also
becomes numerically relevant. In particular, the latter is visible in the
forward-backward asymmetry.

The \SANC Monte Carlo event generator \ReneSANCe and integrator \MCSANCee were used
to produce the numerical results. At the one-loop pure QED level for unpolarized beams, good agreement with the corresponding results of the \BabaYaga code is found. 
The advantages of our codes is implementation of the complete one-loop (electro)weak
corrections and taking into account particle polarizations. 

From Table~\ref{Table:LLAQED5} one can see that the second order ISR corrections are 
numerically relevant for high-precision experiments. 
That brings us to the conclusion that to reduce the theoretical uncertainty we need 
to implement the complete two-loop, i.e., $\mathcal{O}(\alpha^2)$ QED corrections, while 
starting from the third order the corrections can be computed in an approximate manner, i.e.,
with QED showers or even in the collinear LLA approximation.

\section{Funding}
This research was
supported by the Russian Foundation for Basic Research, project
N 20-02-00441.

\section{Acknowledgments}
We are grateful to M.~Potapov 
for the help in preparation of the manuscript.



\begingroup\begin{thebibliography}{10}

\bibitem{Akai:2018mbz}
{SuperKEKB} Collaboration, K.~Akai, K.~Furukawa, and H.~Koiso, {\em Nucl.
  Instrum. Meth. A} {\bf 907} (2018) 188--199,
  \href{http://www.arXiv.org/abs/1809.01958}{{\tt 1809.01958}}.

\bibitem{Asner:2008nq}
D.~M. Asner {\em et al.}, {\em Int. J. Mod. Phys.} {\bf A24} (2009) S1--794,
\href{http://www.arXiv.org/abs/0809.1869}{{\tt 0809.1869}}.

\bibitem{Epifanov:2020elk}
{SCTF} Collaboration, D.~A. Epifanov, {\em Phys. Atom. Nucl.} {\bf 83} (2020),
  no.~6 944--948.

\bibitem{Peng:2020orp}
H.~P. Peng, Y.~H. Zheng, and X.~R. Zhou, {\em Physics} {\bf 49} (2020), no.~8
  513--524.

\bibitem{Belle-II:2018jsg}
{Belle-II} Collaboration, W.~Altmannshofer {\em et al.}, {\em PTEP} {\bf 2019}
  (2019), no.~12 123C01, [Erratum: PTEP 2020, 029201 (2020)],
  \href{http://www.arXiv.org/abs/1808.10567}{{\tt 1808.10567}}.

\bibitem{Roney:2021Bd}
M.~Roney, {\em PoS} {\bf ICHEP2020} (2021) 699.

\bibitem{liptak:ipac2021-thpab022}
Z.~Liptak, M.~Kuriki, and J.~Roney in {\em Proc. IPAC'21}, no.~12,
  pp.~3799--3801, 2021.

\bibitem{CarloniCalame:2000pz}
C.~M. Carloni~Calame, C.~Lunardini, G.~Montagna, O.~Nicrosini, and
  F.~Piccinini, {\em Nucl. Phys. B} {\bf 584} (2000) 459--479,
  \href{http://www.arXiv.org/abs/hep-ph/0003268}{{\tt hep-ph/0003268}}.

\bibitem{CarloniCalame:2001ny}
C.~M. Carloni~Calame, {\em Phys. Lett. B} {\bf 520} (2001) 16--24,
  \href{http://www.arXiv.org/abs/hep-ph/0103117}{{\tt hep-ph/0103117}}.

\bibitem{Balossini:2006wc}
G.~Balossini, C.~M. Carloni~Calame, G.~Montagna, O.~Nicrosini, and
  F.~Piccinini, {\em Nucl. Phys. B} {\bf 758} (2006) 227--253,
  \href{http://www.arXiv.org/abs/hep-ph/0607181}{{\tt hep-ph/0607181}}.

\bibitem{Balossini:2008xr}
G.~Balossini, C.~Bignamini, C.~M.~C. Calame, G.~Montagna, O.~Nicrosini, and
  F.~Piccinini, {\em Phys. Lett. B} {\bf 663} (2008) 209--213,
  \href{http://www.arXiv.org/abs/0801.3360}{{\tt 0801.3360}}.

\bibitem{Jadach:1999vf}
S.~Jadach, B.~Ward, and Z.~Was, {\em Comput. Phys. Commun.} {\bf 130} (2000)
  260--325, \href{http://www.arXiv.org/abs/hep-ph/9912214}{{\tt
  hep-ph/9912214}}.

\bibitem{Jadach:2013aha}
S.~Jadach, B.~F.~L. Ward, and Z.~Was, {\em Phys. Rev. D} {\bf 88} (2013),
  no.~11 114022, \href{http://www.arXiv.org/abs/1307.4037}{{\tt 1307.4037}}.

\bibitem{Arbuzov:2005pt}
A.~B. Arbuzov, G.~V. Fedotovich, F.~V. Ignatov, E.~A. Kuraev, and A.~L.
  Sibidanov, {\em Eur. Phys. J. C} {\bf 46} (2006) 689--703,
  \href{http://www.arXiv.org/abs/hep-ph/0504233}{{\tt hep-ph/0504233}}.

\bibitem{Nugent:2022ayu}
I.~M. Nugent, ``{ee$\in$MC: Simulation of $\bf e^{+}e^{-} \to \mu^{+} \mu^{-}
  (\gamma) $ and $\bf e^{+}e^{-} \to \tau^{+}\tau^{-} (\gamma) $ Events}'',
  preprint (4, 2022), \href{http://www.arXiv.org/abs/2204.02318}{{\tt
  2204.02318}}.

\bibitem{Sadykov:2020any}
R.~Sadykov and V.~Yermolchyk, {\em Comput. Phys. Commun.} {\bf 256} (2020)
  107445, \href{http://www.arXiv.org/abs/2001.10755}{{\tt 2001.10755}}.

\bibitem{Bondarenko:2020hhn}
S.~Bondarenko, Y.~Dydyshka, L.~Kalinovskaya, R.~Sadykov, and V.~Yermolchyk,
  {\em Phys. Rev. D} {\bf 102} (2020), no.~3 033004,
  \href{http://www.arXiv.org/abs/2005.04748}{{\tt 2005.04748}}.

\bibitem{Arbuzov:2021oxs}
A.~B. Arbuzov, S.~G. Bondarenko, L.~V. Kalinovskaya, L.~A. Rumyantsev, and
  V.~L. Yermolchyk, {\em Phys. Rev. D} {\bf 105} (2022), no.~3 033009,
  \href{http://www.arXiv.org/abs/2112.09361}{{\tt 2112.09361}}.

\bibitem{Jegerlehner:2017zsb}
F.~Jegerlehner, {\em EPJ Web Conf.} {\bf 218} (2019) 01003,
  \href{http://www.arXiv.org/abs/1711.06089}{{\tt 1711.06089}}.

\bibitem{Kuraev:1985hb}
E.~A. Kuraev and V.~S. Fadin, {\em Sov. J. Nucl. Phys.} {\bf 41} (1985)
  466--472.

\bibitem{Nicrosini:1986sm}
O.~Nicrosini and L.~Trentadue, {\em Phys. Lett. B} {\bf 196} (1987) 551.

\bibitem{Arbuzov:2021zjc}
A.~Arbuzov, S.~Bondarenko, L.~Kalinovskaya, R.~Sadykov, and V.~Yermolchyk, {\em
  Symmetry} {\bf 13} (2021), no.~7 1256.

\end{thebibliography}\endgroup

\providecommand{\href}[2]{#2}

\end{document}